\documentclass[copyright,creativecommons]{eptcs}
\providecommand{\event}{DCM 2010} % Name of the event you are submitting to
\usepackage{amsfonts}
\usepackage{amsthm}
\usepackage{amssymb}
\usepackage{latexsym}
\usepackage{stmaryrd}
\usepackage{graphicx}
\usepackage{multirow, longtable}
\usepackage{graphicx}
\usepackage{diagrams}
\usepackage{xspace}
\usepackage{rwd-category}
\usepackage{tikz}
\usepackage{tikzquanto}
\usepackage{tikzmcgraph}

\usetikzlibrary{trees}
\usetikzlibrary{topaths}
\usetikzlibrary{decorations.pathmorphing}
\usetikzlibrary{decorations.markings}
\usetikzlibrary{snakes,matrix,backgrounds,folding}
\usetikzlibrary{chains,scopes,positioning,fit}
\usetikzlibrary{arrows,shadows}
\usetikzlibrary{calc} 
\usetikzlibrary{chains}
\usetikzlibrary{shapes,shapes.geometric,shapes.misc}

\newcommand{\mergew}[1]{+_{#1}}
\newcommand{\plugw}[1]{+_{\!#1}^{\!\!\!*}}

\newlength{\hookrightarrowwidth}
\DeclareMathOperator{\matches}{%
\makebox[\hookrightarrowwidth][c]{$\hookrightarrow$}
\hspace*{-\hookrightarrowwidth}%
\makebox[\hookrightarrowwidth][c]{\raise0.5mm\hbox{$\, ^{\sim}$}}%
}

\DeclareMathOperator{\leftmatches}{%
\makebox[\hookrightarrowwidth][c]{$\hookleftarrow$}
\hspace*{-\hookrightarrowwidth}%
\makebox[\hookrightarrowwidth][c]{\raise0.5mm\hbox{$\, ^{\sim}$}}%
}

\DeclareMathOperator{\notleftmatches}{%
\makebox[\hookrightarrowwidth][c]{$\hookleftarrow$}
\hspace*{-\hookrightarrowwidth}%
\makebox[\hookrightarrowwidth][c]{\raise0.5mm\hbox{$\, ^{\sim}$}}%
\hspace*{-\hookrightarrowwidth}%
\makebox[\hookrightarrowwidth][c]{\raise-0.2mm\hbox{\footnotesize{$/$}}}%
}

\newcommand{\seqcompww}[3]{#1\, ;_{#2} #3}

\newcommand{\exten}[2]{{#1}^{\uparrow #2}}

\newcommand{\tensor}{\otimes}
\newcommand{\injmap}{\hookrightarrow}

\newarrow{Partial}----{harpoon}

\newcommand{\catOGraph}{\ensuremath{\textbf{OGraph}}\xspace}
\newcommand{\catGThy}{\ensuremath{\textbf{GThy}}\xspace}

\newcommand{\In}{\textit{In}}
\newcommand{\Out}{\textit{Out}}
\newcommand{\Bound}{\textit{Bound}}
\newcommand{\Isol}{\textit{Isol}}
\newcommand{\Interf}{\textit{Interf}}

\newarrow{Corresponds} <--->

\theoremstyle{plain} % default 
\newtheorem{theorem}{Theorem}[section]
\newtheorem{proposition}[theorem]{Proposition}

\newtheorem{corollary}[theorem]{Corollary}

\theoremstyle{definition}
\newtheorem{definition}[theorem]{Definition}
\newtheorem{definitions}[theorem]{Definitions}

\theoremstyle{remark}
\newtheorem{example}[theorem]{Example}

\title{Open Graphs and Computational Reasoning}
\author{Lucas Dixon
\institute{University of Edinburgh}
\email{l.dixon@ed.ac.uk}
\and
Ross Duncan, Aleks Kissinger
\institute{University of Oxford}
\email{\{ross.duncan, alexander.kissinger\}@comlab.ox.ac.uk}
}
\def\titlerunning{Open Graphs and Computational Reasoning}
\def\authorrunning{L. Dixon, R. Duncan, A. Kissinger}

\begin{document}
%
% \frontmatter          % for the preliminaries
%
%\pagestyle{headings}  % switches on printing of running heads
% \addtocmark{} % additional mark in the TOC

%
% abbreviated author list (for running head)
%\authorrunning{}   
%\tocauthor{  }
%
%\institute{Informatics, University of Edinburgh\\
%  \email{l.dixon@ed.ac.uk}}

%l.dixon@ed.ac.uk 

\maketitle              % typeset the title of the contribution

\begin{abstract}
  We present a form of algebraic reasoning for computational objects
  which are expressed as graphs. Edges describe the flow of data
  between primitive operations which are represented by
  vertices. These graphs have an interface made of half-edges (edges
  which are drawn with an unconnected end) and enjoy rich
  compositional principles by connecting graphs along these
  half-edges. In particular, this allows equations and rewrite rules
  to be specified between graphs.  Particular computational models can
  then be encoded as an axiomatic set of such rules. Further rules can
  be derived graphically and rewriting can be used to simulate the
  dynamics of a computational system, e.g. evaluating a program on an
  input. Examples of models which can be formalised in this way
  include traditional electronic circuits as well as recent
  categorical accounts of quantum information.

% We illustrate the
%  formalism with examples from and note that it also applied .

%A restricted kind
%  of ellipsis notation is then introduced to describe families of
%  graphs and equations.

%  The formalism includes unification of graphs. This supports the
%  analysis of critical pairs and thus Knuth-Bendix style completion
%  for rewrite rules.
  
\end{abstract}

\section{Introduction}

Graphs provide a rich language for specification and reasoning.  Well
known examples include the proof-nets of linear logic
\cite{Girard:proof-nets:96}, Penrose's tensor notation
\cite{Penrose1971Applications-of}, Feynman diagrams, and the common
diagrams used for electronic circuits. Recently, graphs have also been
used to formalise molecular biology~\cite{danos2004formal} and
quantum information processing~\cite{Coecke:2008jo}.

This paper presents \emph{open graphs} as a formal foundation for
reasoning about computational structures.  These graphs have a
directed boundary, visualised as edges entering or leaving the
graph. The boundary of a graph represents the inputs and outputs of
the computation. This lets graphs be interpreted as compound
computations with vertices as the atomic operations. For example, an
electronic circuit that defines the compound logical operation of an
or-gate, using not-gates around an and-gate, can be drawn as:

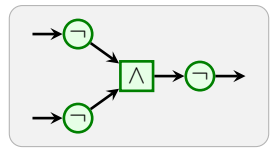
\begin{center}
\begin{tikzpicture}[circuit]
\node (g1) [andg] {$\land$};
\node (x2) [notg, above left=of g1] {$\lnot$};
\node (x1) [notg, below left=of g1] {$\lnot$};
\node (i1) [left=of x1, halfe] {};
\node (i2) [left=of x2, halfe] {};
\node (n3) [notg, right=of g1] {$\lnot$};
\node (o1) [halfe, right=of n3] {};
\path[dline] 
  (i1) edge (x1)
  (i2) edge (x2)
  (x1) edge (g1)
  (x2) edge (g1)
  (g1) edge (n3);
\path[dline] 
  (n3) edge (o1)
;
\end{tikzpicture}
\end{center}
 
The structure and dynamics of a computational model are written in our
formalism as a set of axiomatic rules between graphs. These graphical
rules are declarative in the sense that the graphs they involve can be
rewritten to derive new graphical rules in a conservative manner. In
this way, our formalism is a logical framework to describe models of
computation and derive new results. Another interesting feature of the
formalism is that it provides a clear distinction between sequential
and parallel dynamics in terms of composing rewrites using the
underlying compositional principles of graphs.

Our main contribution is a concise formalism that provides graphs with
a rich compositional structure and a convenient algebraic
language. The formalism includes graphical concepts of addition,
subtraction, tensor product, and substitution, with familiar
laws. Building on these definitions, we develop the mathematics to
support declarative graphical equations and directed rewrites between
graphs. In particular, our development supports rewriting graphs that
contain cycles edges. Care has also been taken to provide a
presentation that can be directly implemented as a software tool. The
intention is to allow techniques from automated reasoning, such as
Knuth-Bendix completation~\cite{knuth-bendix}, to be employed. Thus
graph matching directly incorperates associativity and
commutativity. Another notable feature of the formalism is that it has
a direct correspondence to its visualisation. In this paper, we simply
state the properties of the formalism, leaving the proofs for a longer
manuscript.

For the sake of conciseness and understandability, we illustrate our
formalism with boolean logic circuits. More generally, our formalism
can describe a variety of computational models, including categorical
models of quantum information~\cite{2009:DixonDuncan:AMAI}.

The paper is structured as follows: In~\S\ref{sec:elec}, we introduce
an informal example of reasoning with graphs that model boolean logic
in electronic circuits. This introduces the key challenges. In
\S\ref{sec:related}, we place our contribution in the context of
related work. We introduce our formalism by defining open graphs, in
\S\ref{sec:open-graphs}. We then discuss how these graphs compared and
how one can be found within another in
\S\ref{sec:matching}. Composition of graphs and rewriting are
described in \S\ref{sec:compose-graphs}. In \S\ref{sec:theories} we
introduce the composition of rewrites and how computational models can
be encoded as graphical theories. We conclude and discuss future work
in \S\ref{sec:concl}.

\section{Motivating example: electronic circuits}
\label{sec:elec}

To motivate our formalism, we introduce a graph-based representation
for boolean circuits. We start by giving its generating graphs and
defining axioms. The basic generators for electronic circuits are the
following:

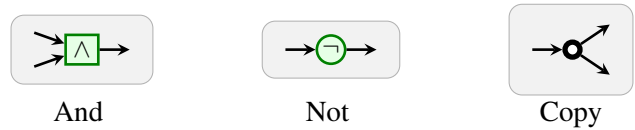
\begin{center}
\begin{tabular}{ccccc}
{
\begin{tikzpicture}[circuit, onegate]
\node (g) [andg] {$\land$};
\node (i1) [halfe, above left=of g.west] {}; 
\node (i2) [halfe, below left=of g.west] {};
\node (o1) [halfe, right=of g] {};
\draw [dline] (i1) -- (g);
\draw [dline] (i2) -- (g);
\draw [dline] (g) -- (o1);
\end{tikzpicture}
}& \hspace{0.5cm} & 
{
\begin{tikzpicture}[circuit]
\node (g) [notg] {$\lnot$};
\node (i1) [halfe, left=of g] {}; 
\node (o1) [halfe, right=of g] {};
\draw [dline] (i1) -- (g);
\draw [dline] (g) -- (o1);
\end{tikzpicture}
}& \hspace{0.5cm} & 
{
\begin{tikzpicture}[circuit]
\node (g) [copyg] {};
\node (i1) [halfe, left=of g] {};
\node (o1) [halfe, above right=of g] {}; 
\node (o2) [halfe, below right=of g] {};
\draw [dline] (i1) -- (g);
\draw [dline] (g) -- (o1);
\draw [dline] (g) -- (o2);
\end{tikzpicture}
} \\
And & & Not & & Copy
\end{tabular}
\end{center}

Any circuit can be built from these components by connecting them
together along the half-edges. The unconnected half-edges that enter a
circuit are the inputs, and those that leave are the outputs. Notice
that copying a value is an explicit operation.

These circuits have two kinds of unitary-generators for the
inputs. There are two unit-generators for the two boolean values and
one counit-generator for when an output is ignored. These are drawn
as:

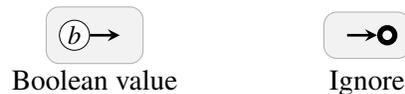
\begin{center}
\begin{tabular}{ccc}
{
\begin{tikzpicture}[circuit]
\node (g) [valg] {$b$};
\node (o1) [halfe, right=of g] {};
\draw [dline] (g) -- (o1);
\end{tikzpicture}
}
&
\hspace{1cm}
& 
{
\begin{tikzpicture}[circuit]
\node (g) [ignoreg] {};
\node (i1) [halfe, left=of g] {};
\draw [dline] (i1) -- (g);
\end{tikzpicture}
}
\\
Boolean value & & Ignore
\end{tabular}
\end{center}

\noindent where we write $b$ for a boolean value which can either be
$F$ for false or $T$ for true.

We can now describe circuits with some values given to them, and with
wires that just stop. This is a simple but important class of
problems. For instance, it includes satisfiability questions, which
are formed by asking whether a given graph can be rewritten to the
single input unit with value $T$. To answer such questions, and more
generally to describe structural equivalences and the dynamics of
boolean circuits, some axioms need to be introduced. For copying and
ignoring values, these are:

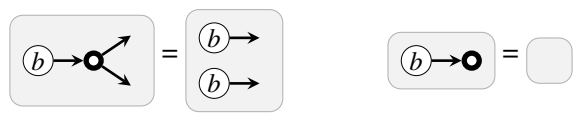
\begin{center}
\mbox{
\begin{tikzpicture}[circuit]
\node (g) [copyg] {};
\node (i1) [valg, left=of g] {$b$};
\node (o1) [halfe, above right=of g] {}; 
\node (o2) [halfe, below right=of g] {};
\draw [dline] (i1) -- (g);
\draw [dline] (g) -- (o1);
\draw [dline] (g) -- (o2);
\end{tikzpicture}
 =
\begin{tikzpicture}[circuit]
\node (i1) [valg] {$b$};
\node (i2) [valg, below=of i1] {$b$};
\node (o1) [halfe, right=of i1] {}; 
\node (o2) [halfe, right=of i2] {};
\draw [dline] (i1) -- (o1);
\draw [dline] (i2) -- (o2);
\end{tikzpicture}
}
\hspace{1cm}
\mbox{
\begin{tikzpicture}[circuit]
\node (i1) [valg] {$b$};
\node (o1) [ignoreg, right=of i1] {};
\draw [dline] (i1) -- (o1);
\end{tikzpicture}
 =
\begin{tikzpicture}[circuit]
\node {\ }; 
\end{tikzpicture}
}
\end{center}

The axioms for conjunction (and-gates: $\land$) and negation
(not-gates: $\lnot$) are:

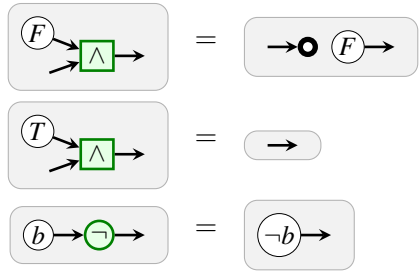
\begin{eqnarray*}
\begin{tikzpicture}[circuit, onegate]
\node (g) [andg] {$\land$};
\node (i1) [valg, above left=of g.west] {$F$}; 
\node (i2) [halfe, below left=of g.west] {};
\node (o1) [halfe, right=of g] {};
\draw [dline] (i1) -- (g);
\draw [dline] (i2) -- (g);
\draw [dline] (g) -- (o1);
\end{tikzpicture}
& = &
\begin{tikzpicture}[circuit, onegate]
\node (v1) [valg] {$F$}; 
\node (o1) [halfe, right=of v1] {};
\node (o2) [ignoreg, left=of v1.center] {};
\node (i2) [halfe, left=of o2] {};
\draw [dline] (v1) -- (o1);
\draw [dline] (i2) -- (o2);
\end{tikzpicture}
\\
\begin{tikzpicture}[circuit,onegate]
\node (g) [andg] {$\land$};
\node (i1) [valg, above left=of g.west] {$T$}; 
\node (i2) [halfe, below left=of g.west] {};
\node (o1) [halfe, right=of g] {};
\draw [dline] (i1) -- (g);
\draw [dline] (i2) -- (g);
\draw [dline] (g) -- (o1);
\end{tikzpicture}
& = &
\begin{tikzpicture}[circuit]
\node (v1) [halfe] {};
\node (o1) [halfe, right=of v1] {};
\draw [dline] (v1) -- (o1);
\end{tikzpicture}
\\
\begin{tikzpicture}[circuit]
\node (v1) [valg] {$b$};
\node (g1) [notg, right=of v1] {$\lnot$}; 
\node (o1) [halfe, right=of g1] {};
\draw [dline] (v1) -- (g1);
\draw [dline] (g1) -- (o1);
\end{tikzpicture}
& = & 
\begin{tikzpicture}[circuit]
\node (v1) [valg] {$\lnot b$};
\node (o1) [halfe, right=of v1] {};
\draw [dline] (v1) -- (o1);
\end{tikzpicture}
\end{eqnarray*}

These provide a sufficient description for evaluating boolean
circuits. Applying these axioms from left to right simulates
evaluation and also performs simplification. The validity of axioms
can be verified by checking the truth tables.

Although the above rules are sufficient for evaluation (when a circuit
has all inputs given), they cannot prove all true equations about
boolean circuits. To get a complete set of equations, new graphical
axioms need to be introduced. For instance, we could easily verify that
\mbox{
\begin{tikzpicture}[circuit]
\node (i1) [halfe] {};
\node (g1) [notg, right=of i1] {$\lnot$}; 
\node (g2) [notg, right=of g1] {$\lnot$}; 
\node (o1) [halfe, right=of g2] {};
\draw [dline] (i1) -- (g1);
\draw [dline] (g1) -- (g2);
\draw [dline] (g2) -- (o1);
\end{tikzpicture}
=
\begin{tikzpicture}[circuit]
\node (i1) [halfe] {};
\node (o1) [halfe, right=of i1] {};
\draw [dline] (i1.center) -- (o1.center);
\end{tikzpicture}
}.

The exhaustive analysis which is performed by examining truth tables
can also be perfomed directly in the graphical language. We can
examine every input to a graphical equation to see if the left and
right hand sides evaluate to the same result. This corresponds to a
proof by exhaustive case analysis. However, rules can also be derived
directly by using the existing equations without examing all
cases. For example, the equations above can be used with the
evaluation axioms to carry out the following derivation:

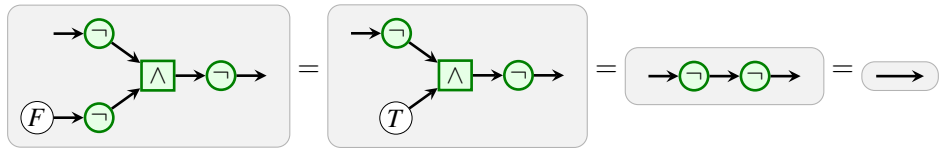
\begin{center}
\begin{tikzpicture}[circuit]
\node (g1) [andg] {$\land$};
\node (x2) [notg, above left=of g1] {$\lnot$};
\node (x1) [notg, below left=of g1] {$\lnot$};
\node (i1) [left=of x1, valg] {$F$};
\node (i2) [left=of x2, halfe] {};
\node (n3) [notg, right=of g1] {$\lnot$};
\node (o1) [halfe, right=of n3] {};
\path[dline] 
  (i1) edge (x1)
  (i2) edge (x2)
  (x1) edge (g1)
  (x2) edge (g1)
  (g1) edge (n3);
\path[dline] 
  (n3) edge (o1)
;
\end{tikzpicture}
$=$
\begin{tikzpicture}[circuit]
\node (g1) [andg] {$\land$};
\node (x2) [notg, above left=of g1] {$\lnot$};
\node (x1) [valg, below left=of g1] {$T$};
\node (i2) [left=of x2, halfe] {};
\node (n3) [notg, right=of g1] {$\lnot$};
\node (o1) [halfe, right=of n3] {};
\path[dline] 
  (i2) edge (x2)
  (x1) edge (g1)
  (x2) edge (g1)
  (g1) edge (n3);
\path[dline] 
  (n3) edge (o1)
;
\end{tikzpicture}
$=$
\begin{tikzpicture}[circuit]
\node (i1) [halfe] {};
\node (g1) [notg, right=of i1] {$\lnot$}; 
\node (g2) [notg, right=of g1] {$\lnot$}; 
\node (o1) [halfe, right=of g2] {};
\draw [dline] (i1) -- (g1);
\draw [dline] (g1) -- (g2);
\draw [dline] (g2) -- (o1);
\end{tikzpicture}
$=$
\begin{tikzpicture}[circuit]
\node (i1) [halfe] {};
\node (o1) [halfe, right=of i1] {};
\path[dline] (i1.center) edge (o1.center);
\end{tikzpicture}
\end{center}

\noindent which proves that giving $F$ to the compound or-gate is the
same as the identiy on the other input. Such derivations can be
exponentially shorter than case-analysis and, in the general setting,
can be carried out in parallel when rewrite rules do not overlap. In
the rest of this paper we focus on a formalism to support this kind of
graphical reasoning.

Another salient feature of graph-based representations is that sharing
and binding can be described using the graph's structure. For example,
consider the following rule: 

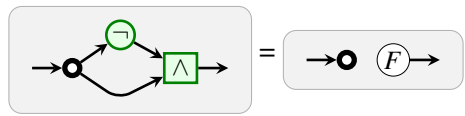
\begin{center}
\mbox{
\begin{tikzpicture}[circuit]
\node (i1) [halfe] {};
\node (g1) [copyg, right=of i1] {};
\node (x2) [halfe, below right=of g1] {};
\node (x1) [notg, above right=of g1] {$\lnot$};
\node (x) [right=of g1] {};
\node (g2) [andg, right=of x] {$\land$};
\node (o1) [halfe, right=of g2] {};
\draw [dline] (i1) -- (g1);
\draw [dline] (g1) -- (x1);
\draw [dline] (x1) -- (g2);
\draw [dline] (g1) -- (x2.center) -- (g2);
\draw [dline] (g2) -- (o1);
\end{tikzpicture}
=
\begin{tikzpicture}[circuit]
\node (i2) [halfe] {};
\node (o2) [ignoreg, right=of i2] {};
\node (v1) [valg, right=of o2.center] {$F$}; 
\node (o1) [halfe, right=of v1] {};
\draw [dline] (v1) -- (o1);
\draw [dline] (i2) -- (o2);
\end{tikzpicture}
}
\end{center}

\noindent The graphical notation treats binding by the structure of
edges. With a formula-based notation this could be described by an
equation between lambda-terms $(\lambda x.\ ((\lnot x) \land x)) =
(\lambda x.\ F)$. This analogy results in composition along half-edges
becoming function application of formula. For example plugging $F$
into the left hand side of the equation could give the formula
$(\lambda x.\ ((\lnot x) \land x)) F$.  Beta-reduction, which reduces
the formula to $(\lnot F) \land F)$, corresponds to the copying
rule. In the graphical language, the beta-reduction step is:

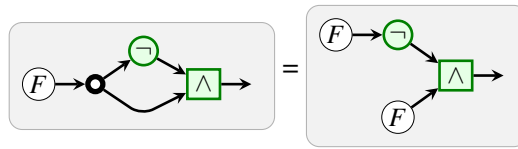
\begin{center}
\mbox{
\begin{tikzpicture}[circuit]
\node (i1) [valg] {$F$}; 
\node (g1) [copyg, right=of i1] {};
\node (x2) [halfe, below right=of g1] {};
\node (x1) [notg, above right=of g1] {$\lnot$};
\node (x) [right=of g1] {};
\node (g2) [andg, right=of x] {$\land$};
\node (o1) [halfe, right=of g2] {};
\draw [dline] (i1) -- (g1);
\draw [dline] (g1) -- (x1);
\draw [dline] (x1) -- (g2);
\draw [dline] (g1) -- (x2.center) -- (g2);
\draw [dline] (g2) -- (o1);
\end{tikzpicture}
=
\begin{tikzpicture}[circuit]
\node (x) {};
\node (x1) [notg, above=of x] {$\lnot$};
\node (i1) [valg, left=of x1] {$F$};
\node (x2) [valg, below=of x] {$F$};
\node (g2) [andg, right=of x] {$\land$};
\node (o1) [halfe, right=of g2] {};
\draw [dline] (i1) -- (x1);
\draw [dline] (x1) -- (g2);
\draw [dline] (x2) -- (g2);
\draw [dline] (g2) -- (o1);
\end{tikzpicture}
}
\end{center}

Notice that the graphical representation controls copying carefully:
by explicit application of equational rules. This is an essential
feature in graphical representations of quantum information, where
copying can only happen in restricted situations. 

These graphs introduce a particular challenge to formalising rewriting
when a direct graph may contain cycles. This is also needed by the
graphical formalisms of quantum information. To understand the
issue, consider the following:

\begin{center}the graph \begin{tikzpicture}[circuit, node distance=0.5em and 0.7em]
\node (x1) [] {};
\node (g) [andg, right=of x1, inner sep=0.4em] {};
\node (x2) [right=of g] {};
\node (x3) [below=of g] {};
\draw [dline] (g) -- (x2.center) -- (x3.center) -- (x1.center) -- (g);
\end{tikzpicture} can be rewritten from left to right by the equation
\begin{tikzpicture}[circuit]
\node (x1) [halfe] {};
\node (g) [andg, right=of x1, inner sep=0.4em] {};
\node (x2) [halfe,right=of g] {};
\draw [dline] (x1) -- (g) -- (x2);
\end{tikzpicture}
=
\begin{tikzpicture}[circuit]
\node (x1) [halfe] {};
\node (x2) [halfe,right=of x1] {};
\draw [dline] (x1.center) -- (x2.center);
\end{tikzpicture}. 
\end{center}

\noindent This results in the graph with a circular edge and no vertices:
\begin{tikzpicture}[circuit, node distance=0.2em and 0.2em]
\node (x1) [] {};
\node (g) [above right=of x1] {};
\node (x2) [below right=of g] {};
\node (x3) [below right=of x1] {};
\draw [dline] (g.center) -- (x2.center) -- (x3.center) -- (x1.center) -- (g.center);
\end{tikzpicture}. 
Such graphs can also be constructed by connecting graphs made only of
edges. These circles in a graph have a natural interpretation for
computational models which interpret graphs be as linear
transformations. These interpretations, which treat spacial-adjacency
as the tensor product, treat circles as scalars. More generally, such
graphs can be understood as traced monoidal categories.

The potential to introduce circles and allow composition along
half-edges is the major challenge for formalisating graph
transformation in this setting. The ability to formally represent such
graphs as finite objects in a computer program, and use them as part
of rewriting, is the main application that our formalism tackles.

\section{Related Work}
\label{sec:related}

%\paragraph{Graph Transformations}

General notions of graph transformations~\cite{Ehrig:Book:2006,
  Baldan-gts}, do not directly provide a suitable basis for formal
equational reasoning about computational objects. In particular,
general graph transformations are not mono, and hence cannot express
equations. Moreoever, the usual formalisms for graphs do not allow
circular edges. Bigraphs provide another general formalism for graph
rewriting, but they are significantly more
complex~\cite{milner2006pure}. Bigraphs use hyper-graphs, where our
edges have a single source and target, and bigraphs also introduce a
rich hierarchical structure.

Rewriting with graph-based presentations of computational systems has
been studied with a variety of formalisms by
Lafont~\cite{Lafont09diagramrewriting,Lafont08,Lafont2003,Lafont1990}. For
instance, Operads and PROPs provide a notable way to rewrite graphical
representations of composable, multi-input and multi-output
functions. A wide variety monoidal categories (and higher-categories)
also enjoy graphical representations~\cite{selinger2009survey}. Open
graphs have a close correspondence to traced symmetric monoidal
categories, but they absorb the braiding operations on tensor
products. The main difference with our work is that we formalise
graphs directly, rather than treat them as a presentation of an
algebraic structure. This makes our formalism particularly amenable to
software implementation~\footnote{For example, see our implementation
  of Quantomatic:
  \url{http://dream.inf.ed.ac.uk/projects/quantomatic}. }. Our notion
of matching directly absorbs associative-commutative structure.

Work in systems biology provides another formalism for graphs that is
similar to the one presented here~\cite{danos2004formal}. The main
difference is that our notion of matching is stricter: we require all
incident edges to be specified in a rule. This is needed to ensure
that a vertex has a fixed number of inputs and outputs.

In~\cite{2009:DixonDuncan:AMAI} we introduced a formalism for
reasoning about categorical models of quantum information. We have
since made several important improvements: matching and composition
are now dual notions which allows a concise definition rewriting.

% add: Corradini: involving 2-categories 
% add: operads and/or PROPs they are rather standard tools for formalizing
%rewriting techniques. See: 
%  http://dx.doi.org/10.2991/jnmp.2006.13.s.8
%  http://iml.univ-mrs.fr/~lafont/pub/diagrams.pdf
%  http://www.maths.gla.ac.uk/~tl/book.html
%
%%%% DONE

% say something about loopy stuff; From reviewer: I would also have
%liked a comment about iteration and recursion (ie, cyclic graphs)
%earlier than the final sentence.  You start off talking about ``flow''
%of data, and your rewriting semantics assumes this flow is in some
%sense instantaneous.  So how does (will) this impact cyclic
%structures? 
%%%% DONE

%%%%%%%%%%%%%%%%%%%%%%%%%%%%%%%%%%%%%%%%%%%%%%%%%%%%%%%%%%%%%%%%%%%%%%%%%%%%%%%%%%%%%%%%%%%%%%%%%%%%

\section{Open Graphs and Embeddings}
\label{sec:open-graphs}
%% Lucas

\begin{definition}[Pre-Open graph]
Let $\mathcal G$ be the following finite coproduct sketch.
\begin{diagram}
 E &  \pile{\rTo^{s} \\ \rTo_{t}} & V + \epsilon \\
\end{diagram}

We call models of $\mathcal G$ in $\catSet$ \emph{pre-open graphs}.
\end{definition}

For a pre-open graph $G$, $G(V + \epsilon)$ identifies a set of
points, with $V$ being the vertices, and $\epsilon$ being points on an
edge, which we call \emph{edge points}. If $G(\epsilon) = \{ \}$, we
recover the usual notion of directed graph. Edge-points provide a
combinatorial description of `dangling' edges, `half' edges, and edges
attached to themselves (circles). In particular, they allow graphs to
be composed by edge-points. Where convenient, we will use subscript
notation to refer to elements in the model of $\mathcal G$, i.e. $s_G
:= G(s)$ and $E_G := G(E)$.

A pre-open graph is called an \emph{open graph} if the edge points
behave as if they were really points occurring within a single
directed edge.

% \begin{definition}[Open graph]
% 	\label{def:open-graph}
%   For a pre-open graph $G$, let $s' : E_s \rightarrow \epsilon$, $t' : E_t \rightarrow \epsilon$ be the restrictions of $s$ and $t$ to edge points. $G$ is called an \emph{open graph} when:
% \begin{enumerate}
% \item $s'$ and $t'$ are injective, and
% \item $E_s$ and $E_t$ cover $\epsilon$, i.e. $s'(E_s) \cup t'(E_t) = \epsilon$.
% \end{enumerate} 
% \end{definition}

\begin{definition}[Open graph]
	\label{def:open-graph}
  For a pre-open graph $G$, let $s' : E_s \rightarrow \epsilon$, $t' :
  E_t \rightarrow \epsilon$ be the restrictions of $s$ and $t$ to edge
  points. $G$ is called an \emph{open graph} when $s'$ and $t'$ are
  injective. Let $\catOGraph$ be the full subcategory of $Mod(\mathcal
  G)$ whose objects are open graphs.
\end{definition}

Injectivity ensures edges are not allowed to branch. Each point has at
most one incoming edge (called an \emph{in-edge}) and one outgoing
edge (called an \emph{out-edge}). Only vertices may have multiple
in-edges and/or multiple out-edges.

\begin{definitions}
If a point $p \in \epsilon_G$ has one out-edge, but no in-edges, it is
called an \emph{input}. Similarly, a point with one in-edge and no
out-edges is called an \emph{output}. The inputs and outputs define a
graph's \emph{boundary}. If a point has no in-edges and no out-edges,
it is called an \emph{isolated point}. We use the following notation
for these subsets of $\epsilon_G$:
\begin{enumerate}
	\item $\In(G)$ the set of inputs,
	\item $\Out(G)$ the set of outputs,
	\item $\Bound(G) = \In(G) \cup \Out(G)$ the set of boundary points, and
	\item $\Isol(G)$ the set of isolated points.
\end{enumerate}
A graph consisting of only isolated points is called a
\emph{point-graph}. These will be used to define how graphs are
composed.
\end{definitions}

\begin{example} 
\label{ex:ograph} 
The following is an illustration of an open graph.

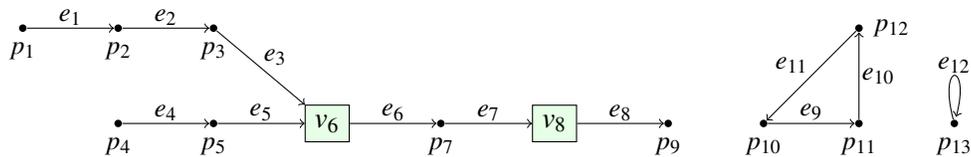
\begin{center}
  \begin{tikzpicture}[mcgraph, node distance=3em and 3em]
  \node[ipoint] (p1) {};  
	\node[ipointlabel,below=of p1] (p1l) {$p_1$};

  \node[right=of p1,ipoint] (p2) {};  
  \node[ipointlabel,below=of p2] (p2l) {$p_2$};  

  \node[right=of p2,ipoint] (p3) {}; 
  \node[ipointlabel,below=of p3] (p3l) {$p_3$};  

  \node[ipoint,below=of p2] (p4) {};  
	\node[ipointlabel,below=of p4] (p4l) {$p_4$}; 

  \node[right=of p4,ipoint] (p5) {};  
  \node[ipointlabel,below=of p5] (p5l) {$p_5$};  

  \node[right=of p5,vpoint] (v6) {$v_6$};
  \node[right=of v6,ipoint] (p7) {};
  \node[ipointlabel,below=of p7] (p7l) {$p_7$};   
  \node[right=of p7,vpoint] (v7) {$v_8$};  

  \node[right=of v7,ipoint] (p8) {};
	\node[ipointlabel,below=of p8] (p8l) {$p_9$};  

  \path[->,elabel,above] (p1) edge node {$e_1$} (p2)
            (p2) edge node {$e_2$} (p3)
            (p3) edge node[above right] {$e_3$} (v6)

            (p4) edge node {$e_4$} (p5)
            (p5) edge node {$e_5$} (v6)

            (v6) edge node {$e_6$} (p7)
						(p7) edge node {$e_7$} (v7)
            (v7) edge node {$e_8$} (p8);

  \node[right=of p8,ipoint] (p9) {};  
  \node[ipointlabel,below=of p9] (p9l) {$p_{10}$};  
  \node[right=of p9,ipoint] (p10) {};  
  \node[ipointlabel,below=of p10] (p10l) {$p_{11}$};  
  \node[above=of p10,ipoint] (p11) {};  
  \node[ipointlabel,right=of p11] (p11l) {$p_{12}$};  

  \node[right=of p10,ipoint] (p12) {};
	\node[ipointlabel,below=of p12] (p12l) {$p_{13}$};  

  \path[->,elabel] 
            (p9) edge node[above] {$e_9$} (p10)
            (p10) edge node[right] {$e_{10}$} (p11)
            (p11) edge node[above left] {$e_{11}$} (p9)
            (p12) edge[loop above] node[above] {$e_{12}$} ();
  \end{tikzpicture}
\end{center}
where $p_i$ are edge points and $v_i$ are vertices. Note that $p_1$, $p_4$ and $p_8$ are boundary points.

The following is not an open graph, because the map $s$ is not
injective for $p_{15}$ and $t$ is not injective for $p_{15}$ and $p_{16}$ 
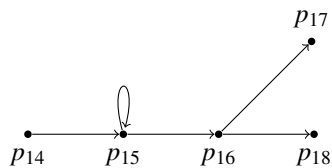
\begin{center}
\begin{tikzpicture}[mcgraph, node distance=3em and 3em]
  \node[ipoint] (p14) {};  
  \node[ipointlabel,below=of p14] (p14l) {$p_{14}$};  

  \node[right=of p14,ipoint] (p15) {};  
  \node[ipointlabel,below=of p15] (p15l) {$p_{15}$};  

  \node[right=of p15,ipoint] (p16) {};  
  \node[ipointlabel,below=of p16] (p16l) {$p_{16}$};  

  \node[above right=of p16,ipoint] (p17) {};  
  \node[ipointlabel,above=of p17] (p17l) {$p_{17}$};  

  \node[right=of p16,ipoint] (p18) {};  
  \node[ipointlabel,below=of p18] (p18l) {$p_{18}$};  

  \path[->,elabel] 
  (p14) edge (p15)
  (p15) edge[loop above]  ()
  (p15) edge (p16)
  (p16) edge (p17)
  (p16) edge (p18)
  ;
\end{tikzpicture}
\end{center}
\end{example}

% In order to obtain a more topological notion of graph embedding, well-suited to matching and graph rewriting, we will define the following notion.

\begin{definition}
	An \emph{open embedding} is a monomorphism $e : G \rightarrow H$
  between pre-open graphs that is full on vertices: for all $p$ if
  $e(p) \in V_H$ then all edges adjacent to $e(p)$ are in the image of
  $e$.
\end{definition}

Open embeddings describe how one graph can be found within another one
and they will be used to identify the parts of a graph which can be
rewritten by a rule.  The identity map, the empty map, and graph
isomorphisms are open embeddings.  

\begin{example}
	The graph on the left has an open embedding into example \ref{ex:ograph}, whereas the graph on the right does not.
	
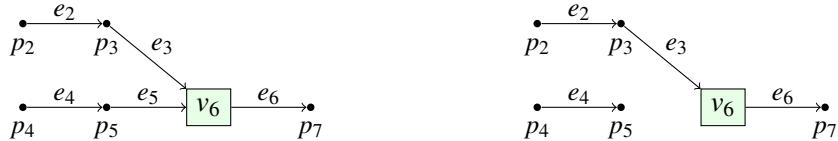
\begin{center}
	\begin{tikzpicture}
		\node[right=of p1,ipoint] (p2) {};  
	  \node[ipointlabel,below=of p2] (p2l) {$p_2$};  

	  \node[right=of p2,ipoint] (p3) {}; 
	  \node[ipointlabel,below=of p3] (p3l) {$p_3$};  

	  \node[ipoint,below=of p2] (p4) {};  
		\node[ipointlabel,below=of p4] (p4l) {$p_4$}; 

	  \node[right=of p4,ipoint] (p5) {};  
	  \node[ipointlabel,below=of p5] (p5l) {$p_5$};  

	  \node[right=of p5,vpoint] (v6) {$v_6$};
	  \node[right=of v6,ipoint] (p7) {};
	  \node[ipointlabel,below=of p7] (p7l) {$p_7$};
	
		\path[->,elabel,above]
	            (p2) edge node {$e_2$} (p3)
	            (p3) edge node[above right] {$e_3$} (v6)

	            (p4) edge node {$e_4$} (p5)
	            (p5) edge node {$e_5$} (v6)

	            (v6) edge node {$e_6$} (p7);
	\end{tikzpicture}
	\qquad\qquad\qquad
	\begin{tikzpicture}
		\node[right=of p1,ipoint] (p2) {};  
	  \node[ipointlabel,below=of p2] (p2l) {$p_2$};  

	  \node[right=of p2,ipoint] (p3) {}; 
	  \node[ipointlabel,below=of p3] (p3l) {$p_3$};  

	  \node[ipoint,below=of p2] (p4) {};  
		\node[ipointlabel,below=of p4] (p4l) {$p_4$}; 

	  \node[right=of p4,ipoint] (p5) {};  
	  \node[ipointlabel,below=of p5] (p5l) {$p_5$};  

	  \node[right=of p5,vpoint] (v6) {$v_6$};
	  \node[right=of v6,ipoint] (p7) {};
	  \node[ipointlabel,below=of p7] (p7l) {$p_7$};
	
		\path[->,elabel,above]
	            (p2) edge node {$e_2$} (p3)
	            (p3) edge node[above right] {$e_3$} (v6)

	            (p4) edge node {$e_4$} (p5)

	            (v6) edge node {$e_6$} (p7);
	\end{tikzpicture}
	\end{center}
\end{example}

The set of all open embeddings into an open graph, $G$, is analogous
to the notion of open subsets of $G$, if $G$ were to be considered a
graph in the topological sense. The notion of open is a midway point
between the usual combinatorial definition, and the idea of a graph as
a topological space made by gluing together 1-dimensional manifolds.

%Edge-points permits relatively simple definitions of composition, as
%we will see later.

\section{Homeomorphism and Matching}
\label{sec:matching}

While edge points are a useful tool for composition, the number of
edge points along an edge is irrelevant to the intended meaning of the
graph. Just as two copies of the interval $[0,1]$ glued end to end are
homeomorphic to just one copy, we shall define a concept of
homeomorphic open graphs which is a course-graining of graph
isomorphism.

\begin{definition}
  We say that $G$ \emph{contracts} to $H$, if $G$ can be made isomorphic with $H$ by replacing any number of subgraphs of this form:
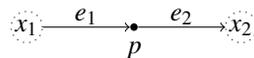
\begin{center}
  \begin{tikzpicture}[mcgraph, node distance=3em and 3em]
  \node[bndry] (p1) {$x_1$}; 

  \node[right=of p1,ipoint] (p2) {};  
  \node[ipointlabel,below=of p2] (p2l) {$p$};  

  \node[right=of p2,bndry] (p3) {$x_2$};

  \path[->,elabel,above] (p1) edge node {$e_1$} (p2)
            (p2) edge node {$e_2$} (p3) ;
\end{tikzpicture}
\end{center}
where $p \in \epsilon$ and $x_1, x_2$ are (not necessary distinct) points in  $V + \epsilon$, with a graphs of this form:
\begin{center}
  \begin{tikzpicture}[mcgraph, node distance=3em and 3em]
    \node[bndry] (p1) {$x_1$};
    \node[right=of p1,bndry] (p3) {$x_2$};

    \path[->,elabel,above] (p1) edge node {$e_3$} (p3);
  \end{tikzpicture}
\end{center}
In such a case, we write $H \preceq G$. Let $\sim$ then be the symmetric closure of $\preceq$. We say two graphs are \emph{homeomorphic} iff $G \sim H$.
\end{definition}

\begin{proposition}\label{pro:homeo}
	$\preceq$ is a well-founded, confluent partial order, up to graph isomorphism.
\end{proposition}

\begin{definition}\label{def:proper}
	An open graph is said to be \emph{proper} if it is minimal with respect to $\preceq$.
\end{definition}

\begin{proposition}
  Every graph has a unique minimal form. For graphs $G$, $H$, $G \sim
  H$ iff their minimal forms are isomorphic.
\end{proposition}

Proper graphs are of particular interest for computing with open
graphs because they provide a computable, minimal representations for
open graphs.

\begin{example} 
\label{ex:open-graph}
The associated proper open graph of example~\ref{ex:ograph} is:

\begin{center}
  \begin{tikzpicture}[mcgraph, node distance=3em and 5em]
  \node[ipoint] (p1) {}; 
	\node[ipointlabel,below=of p1] {$p_1$};
  \node[below=of p1,ipoint] (p4) {};
	\node[ipointlabel,below=of p4] {$p_4$};
  \node[right=of p4,vpoint] (v6) {$v_6$};  
  \node[right=of v6,vpoint] (v7) {$v_7$};  

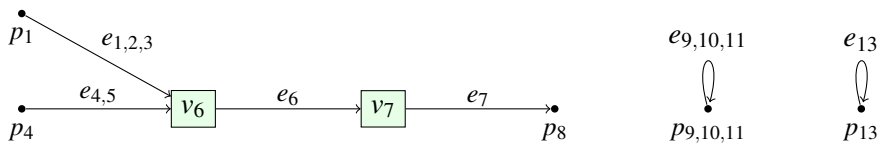
  \node[right=of v7,ipoint] (p8) {};  
	\node[ipointlabel,below=of p8] {$p_8$};
	
  \node[right=of p8,ipoint] (p9) {};
	\node[ipointlabel,below=of p9] {$p_{9,10,11}$};

  \node[right=of p9,ipoint] (c13) {};
	\node[ipointlabel,below=of c13] {$p_{13}$};

  \path[->,elabel,above] (p1) edge node[above right] {$e_{1,2,3}$} (v6)
            (p4) edge node[above] {$e_{4,5}$} (v6)
            (v6) edge node {$e_6$} (v7)
            (v7) edge node {$e_7$} (p8)
            ;

  \path[->] (p9) edge[loop above] node[above] {$e_{9,10,11}$} ()
            (c13) edge[loop above] node[above] {$e_{13}$} ();

  \end{tikzpicture}
\end{center}
\end{example}

By considering open embeddings up to the relation $\sim$ we get a more
general notion of what it means for a graph to be ``inside'' of
another. This abstracts over any intermediate edge-points.

\begin{definition}\label{def:matching}
	A graph $L$ is said to \emph{match} $G$, denoted by $L \matches
  G$, if there exists some $G' \sim G$ and an open embedding $e : L
  \rightarrow G'$. Such an embedding is called a \emph{matching} of
  $L$ on $G$.
\end{definition}

\begin{example} 
\label{ex1graph}
Let $G$ be the open graph:

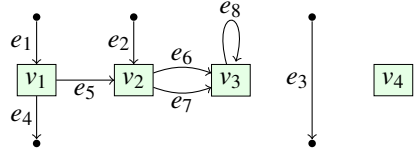
\begin{center}
  \begin{tikzpicture}[mcgraph]
  \matrix (m) [matrix of math nodes,row sep=1.5em, column sep=2em] { 
   \node[ipoint] (s1) {};  &  \node[ipoint] (s2) {};  & & \node[ipoint] (s3) {}; &    \\
   \node[ivert] (v1) {v_1};  &   \node[ivert] (v2) {v_2};     &  \node[ivert] (v3) {v_3};   &       &  \node[ivert] (v4) {v_4};   \\
   \node[ipoint] (t1) {};   &      &    &  \node[ipoint] (t2) {};&  \\
  };

	% \node [ipointlabel,above=of s1] {$s_1$};
	% \node [ipointlabel,above=of s2] {$s_2$};
	% \node [ipointlabel,below=of t1] {$t_1$};
	% \node [ipointlabel,below=of t2] {$t_2$};
	% \node [ipointlabel,above=of s3] {$s_3$};

  \path[->] (s1) edge node[left,elabel] {$e_1$} (v1) 
            (s2) edge node[left,elabel] {$e_2$} (v2)
            (s3) edge node[left,elabel] {$e_3$} (t2); 

  \path[->] (v1) edge node[left,elabel] {$e_4$} (t1) 
            (v1) edge node[below,elabel] {$e_5$} (v2)
            (v2) edge[bend left=20] node[above,elabel] {$e_6$} (v3)
            (v2) edge[bend right=20] node[below,elabel] {$e_7$} (v3);

  \path[->] (v3) edge[loop above] node[above,elabel] {$e_8$} (e8);

%  \path[->] (m-3-5) edge[loop above] node[above] {$e_8$} (m-3-5);
  
\end{tikzpicture}
\end{center}

\noindent then the following graph, $H$, matches $G$, where each edge
of the form $e^{(n)}_i$ in $H$ maps on to an edge expanded from the
edge $e_i$ in $G$.

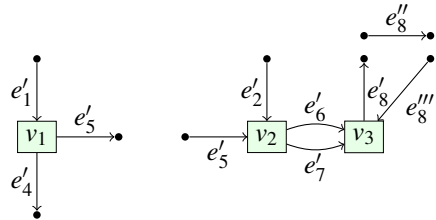
\begin{center}
  \begin{tikzpicture}[mcgraph, node distance=2em and 2em]
	\node[ipoint] (s1) {};
	\node[ivert,below=of s1] (v1) {$v_1$};
	\node[ipoint,below=of v1] (t1) {};
	\node[ipoint,right=of v1] (s1v2) {};
	
	\node[ipoint,right=of s1v2] (s1v1) {};
	\node[ivert,right=of s1v1] (v2) {$v_2$}; 
	\node[ivert,right=of v2] (v3) {$v_3$};
	
	\node[ipoint,above=of v2] (s2) {};
	
	\node[ipoint,above=of v3] (s1v3) {};
	\node[ipoint,right=of s1v3] (s4v3) {};
	
	\node[ipoint,node distance=0.5em and 0.5em,above=of s1v3] (s2v3) {};
	\node[ipoint,right=of s2v3] (s3v3) {};

	%   \node [ipointlabel, above=of s1] {$s_1$};
	%   \node [ipointlabel, above=of s2] {$s_2$};
	%   \node [ipointlabel, below=of t1] {$t_1$};
	% \node [ipointlabel, below=of s1v1] {$t_2$};
	% \node [ipointlabel, below=of s1v2] {$t_3$};
	% \node [ipointlabel, above=of s2v3] {$v_3$};
	% \node [ipointlabel, above=of s3v3] {$v_3$};

  \path[->] (s1) edge node[left,elabel] {$e_1'$} (v1) 
            (s2) edge node[left,elabel] {$e_2'$} (v2);

  \path[->] (v1) edge node[left,elabel] {$e_4'$} (t1) 
            (v1) edge node[above,elabel] {$e_5'$} (s1v2)
            (s1v1) edge node[below,elabel] {$e_5'$} (v2)
            (v3) edge node[right,elabel] {$e_8'$} (s1v3)
            (s2v3) edge node[above,elabel] {$e_8''$} (s3v3)
            (s4v3) edge node[below right,elabel] {$e_8'''$} (v3)
            (v2) edge[bend left=20] node[above,elabel] {$e_6'$} (v3)
            (v2) edge[bend right=20] node[below,elabel] {$e_7'$} (v3);
%  \path[->] (m-3-5) edge[loop above] node[above] {$e_8$} (m-3-5);
\end{tikzpicture}
\end{center}
\end{example}

\begin{example} 
\label{ex3} 
Some more examples of graphs and matchings. The graph in the centre
matches those on its right, but not those on its left:
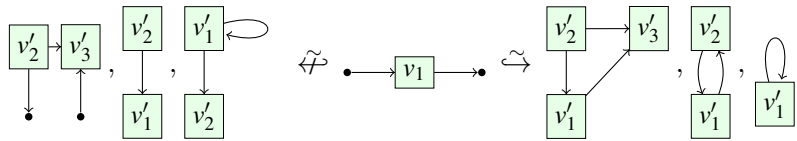
\begin{center}
%\begin{tabular}{cccccccc}
  \begin{tikzpicture}[mcgraph]
  \node[ipoint] (t1) {}; 
  \node[ipoint, right=of t1] (s1) {}; 
  \node[above=of t1,ivert] (v2) {$v_2'$};  
  \node[above=of s1,ivert] (v3) {$v_3'$};  
  \path[->] (v2) edge (v3)
            (v2) edge (t1)
            (s1) edge (v3);
  \end{tikzpicture}
  $,$ 
  \begin{tikzpicture}[mcgraph]
  \node[ivert] (v1) {$v_1'$}; 
  \node[above=of v1, ivert] (s1) {$v_2'$};  
  \path[->] (s1) edge (v1);
  \end{tikzpicture}
  $,$
  \begin{tikzpicture}[mcgraph]
  \node[ivert] (v1) {$v_1'$}; 
  \node[ivert, below=of v1] (v2) {$v_2'$}; 
  \path[->] (v1) edge (v2);
  \path[->] (v1) edge[loop right] (e8);
  \end{tikzpicture}
   $\ \notleftmatches\ $ 
  \begin{tikzpicture}[mcgraph]
  \node[ivert] (v1) {$v_1$}; 
  \node[left=of v1, ipoint] (s1) {};  
  \node[right=of v1,ipoint] (t1) {};  
  \path[->] (v1) edge (t1)
            (s1) edge (v1);
  \end{tikzpicture}
  \ $\matches$\ \ 
  \begin{tikzpicture}[mcgraph]
  \node[ivert] (v1) {$v_1'$}; 
  \node[above=of v1, ivert] (s1) {$v_2'$};  
  \node[above right=of v1,ivert] (t1) {$v_3'$};  
  \path[->] (v1) edge (t1)
            (s1) edge (t1)
            (s1) edge (v1);
  \end{tikzpicture}
  $,$
  \begin{tikzpicture}[mcgraph]
  \node[ivert] (v1) {$v_1'$}; 
  \node[above=of v1, ivert] (s1) {$v_2'$};  
  \path[->] (v1) edge [bend right=20] (s1)
            (s1) edge [bend right=20] (v1);
  \end{tikzpicture}
  $,$
  \begin{tikzpicture}[mcgraph]
  \node[ivert] (v1) {$v_1'$}; 
  \path[->] (v1) edge[loop above] node[above] {} (e8);
  \end{tikzpicture}
  % \\
  % & & \\
  % \begin{tikzpicture}[mcgraph]
  % \node[ivert] (v1) {$v_1$}; 
  % \node[above left=of v1, bndry] (s1) {$s_1$};  
  % \node[above right=of v1,bndry] (t1) {$t_1$};  
  % \path[->] (v1) edge (t1)
  %           (s1) edge (v1);
  % \end{tikzpicture}
%\end{tabular}
\end{center}
\end{example}

%%%%%%%%%%%%%%%%%%%%%%%%%%%%%%%%%%%%%%%%%%%%%%%%%%%%%%%%%%%%%%%%%%%%%%%%%%%%%%%%%%%%%%%%%%%%%%%%%%%%
\section{Composition and Rewriting}
\label{sec:compose-graphs}

In this section, we define the notions of composing graphs along their
boundaries and performing rewrites of graphs embedded inside of other
graphs. First, we define a notion of boundary and interface for a
graph.

\begin{definition}[Boundary embedding]
  A \emph{boundary embedding} $b$ is an open embedding from a
  point graph, $P$, to a graph, $G$, such that only boundary
  points are in the image of $b$: $\forall x \in \epsilon_P.\ b(x)
  \in \Bound(G)$.
\end{definition}

Boundary embeddings identify a subset of a graph's boundary points
along which the graph can be composed with another graph. In
particular, they let us define the \emph{interface} of a graph.

\begin{definition}[Interface]
  The \emph{interface} of a graph $G$, written $\Interf(G)$, is a
  tuple $(P_I,b_I,P_O,b_O)$, where $P_I$ and $P_O$ are point graphs
  such that $b_I : P_I \rightarrow G$ is a boundary embedding that is
  surjective on $\In(G)$, and $b_O : P_O \rightarrow G$ is a boundary
  embedding that is surjective $\Out(G)$.
\end{definition}

\begin{definition}
  Two graphs $G$ and $G'$ are said to have \emph{the same boundary}
  (or \emph{the same interface}) when $\Interf(G)$ and $\Interf(G')$ are
  isomorphic.
\end{definition}

% \begin{definition}[Complementary Boundary Embeddings]
%   A pair of boundary embeddings $b : P \rightarrow G$ and $b' : P
%   \rightarrow G'$, are called \emph{complementary} when one takes
%   point to an input iff the other takes the point to an
%   output: $$\forall x \in \epsilon_P.\ b(x) \in s^{\rightarrow}_G
%   \Leftrightarrow b'(x) \in t^{\rightarrow}G'$$
% \end{definition}
% 
% Complementary boundary embeddings identify boundary points in two
% graphs that agree in direction so as to allow the graphs to be
% composed at these boundary points.

Colimits can be used to `glue' multiple graphs together. Informally,
they define minimal graphs that a given set of other graphs can be
embedded into. We will define \emph{open colimits} as colimits with
conditions to ensure that they merge graphs coherently. This will
provide a notion of merged graph that is coherent with its components,
and in particular, with which we can define composition along
half-edges. Whereas the category $Mod(\mathcal G)$ has all small
colimits, $\catOGraph$ only has certain small colimits, namely those
that will not break the injectivity condition on the edge points.

\begin{definition}
	A pair of arrows $f : G \rightarrow H_1$, $g : G \rightarrow H_2$ in $\catOGraph$ are said to be \emph{boundary-coherent} when
	\begin{enumerate}
		\item for all $p \in \Bound(G)$, $f(p) \notin \Bound(H_1) \implies g(p) \in \Bound(H_2)$, and
		\item for all $p \in \Isol(G)$, either $f(p)$ is an isolated point, $g(p)$ is an isolated point, or $f(p)$ and $g(p)$ are compatible boundaries. That is, one is an input iff the other is an output.
	\end{enumerate}
\end{definition}

\begin{definition}
  A colimit of a diagram $D$ in $\catOGraph$ is called an \emph{open colimit}
  when all pairs of arrows in $D$ are boundary-coherent.
\end{definition}

%the special cases of open
%coequalisers, sums (which are automatically open), and 

We will focus on the notion of open pushouts, which we shall call
\emph{mergings}, that arrises from open colimits. In particular, given
graphs $G_1$, $G_2$ and $K$, with open embeddings $e_1 : K \injmap G_1$ and
$e_2 : K \injmap G_2$, a graph $M$ is called the \emph{merging} of $G_1$
and $G_2$ (on $K$ by $e_1$ and $e_2$) when it is defined by the pushout:
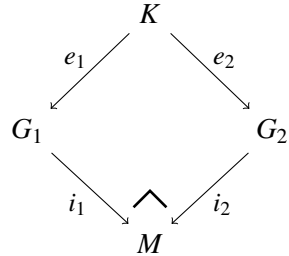
\begin{center}
  \begin{tikzpicture}[text height=1.5ex]
	  \node[] (top) {$K$}; 
	  \node[below left=of top] (left) {$G_1$}; 
	  \node[below right=of top] (right) {$G_2$}; 
	  \node[below=of top] (mid) {}; 
	  \node[below=of mid] (bot) {$M$};
	  \path[->] (top) edge node [above left,elabel] {$e_1$} (left)
	            (top) edge node [above right,elabel] {$e_2$} (right)
	            (left) edge node [below left,elabel] {$i_1$} (bot)
	            (right) edge node [below right,elabel] {$i_2$} (bot)
	            ;
 
	  { [node distance=0em and 0em]
	     \node[above=of bot, circle, inner sep=2pt] (pushout) {};
	     \draw[line width=1pt] (pushout.west) -- (pushout.north) -- (pushout.east); 
	  }
	\end{tikzpicture}
\end{center}
\noindent This makes $M$ the smallest graph containing $G_1$ and $G_2$
merged exactly on the graph $K$, as identified by $e_1$ and $e_2$. Given
$p := (K, e_1, e_2)$, we use the notation $G_1 \mergew{p} G_2$ to
denote $M$. If the open embeddings associated with pushout triples $p$
and $q$ are disjoint, we can take the merging to be strictly
associative and omit parentheses:
$G_1 \mergew{p} G_2 \mergew{q} G_3 :=
(G_1 \mergew{p} G_2) \mergew{q} G_3 =
G_1 \mergew{p} (G_2 \mergew{q} G_3)
$

%% *** actually we can say more than this: even if not disjoint,
%% there's an isomorphism between them. 

%\medskip

\begin{theorem} 
  \label{thm:ograph-colimits}
	$\catOGraph$ has all small open colimits. Furthermore, colimit maps are full on vertices.
\end{theorem}
%%% see scraps/proof.tex for the proof; update the proof if you change the def!

\begin{corollary}
	Given a merging,
	
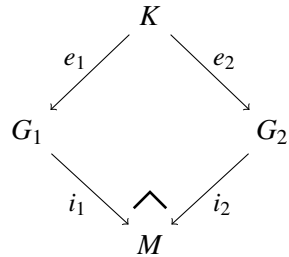
\begin{center}
	  \begin{tikzpicture}[text height=1.5ex]
	  \node[] (top) {$K$}; 
	  \node[below left=of top] (left) {$G_1$}; 
	  \node[below right=of top] (right) {$G_2$}; 
	  \node[below=of top] (mid) {}; 
	  \node[below=of mid] (bot) {$M$};
	  \path[->] (top) edge node [above left,elabel] {$e_1$} (left)
	            (top) edge node [above right,elabel] {$e_2$} (right)
	            (left) edge node [below left,elabel] {$i_1$} (bot)
	            (right) edge node [below right,elabel] {$i_2$} (bot)
	            ;

	  { [node distance=0em and 0em]
	     \node[above=of bot, circle, inner sep=2pt] (pushout) {};
	     \draw[line width=1pt] (pushout.west) -- (pushout.north) -- (pushout.east); 
	  }
	\end{tikzpicture}
	\end{center}
	the maps $i_1$, $i_2$ are open embeddings.
\end{corollary}
%%% has proof in scraps/proof.tex

This ensures that the visual intuition of finding one graph in another
is preserved: merging graphs does not change their internal structure
and hence we can find a graph $G$ (an open embedding of it) within a
any merging of $G$ with any other graph.

\begin{definition}[Tensor]
  When the graph being merged-on is empty, we call the resulting
  merged graph, $G \mergew{(\emptyset,\emptyset,\emptyset)} H$, the
  \emph{tensor product}, and write it as $G \tensor H$.
\end{definition}

Tensor composition corresponds to placing graphs side by side. To
define composition along half-edges, we use point graphs and a special
kind of open embedding that identifies (part of) the boundary.

\begin{definition}[Plugging]
  A graph merging $G \mergew{(P,b,b')} G'$ is called a
  \emph{plugging}, and is written $G \plugw{(P,b,b')} G'$, when $P$
  is a point graph and $b$ and $b'$ are boundary-coherent
  boundary embeddings.
\end{definition}

\begin{example} 
\label{ex4:plug} 
An example of plugging using pushouts. The grey boxes are drawn around
the graphs involved to distinguish between edges in the graphs and
those of the pushout diagram.

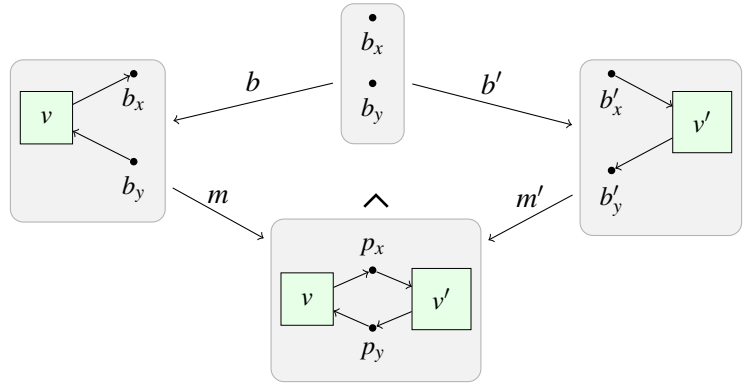
\begin{center}
  \begin{tikzpicture}[mcgraph,node distance=5em and 7em]
  \node[ipoint] (js1) {};  
  \node[ipointlabel, below=of js1] (js1l) {$b_{x}$};  

  { [node distance=0.5em and 2em]
  \node[ipoint, below=of js1l] (js2) {};  
  \node[ipointlabel, below=of js2] (js2l) {$b_{y}$};  
%  \path[->] (js1) edge (jt1)
%            (js2) edge (jt2);
  }

  \begin{pgfonlayer}{background}
  \node[rectangle, fill=black!5, draw=black!30,rounded corners=1ex, fit=(js1) (js2) (js2l) (js1l)] (J) {};  
  \end{pgfonlayer}

  \node[ipoint, left=of J] (lt1) {};
  \node[ipointlabel, below=of lt1] (lt1l) {$b_x$};  
  { [node distance=0.5em and 2em]
  \node[ivert, inner sep=0.7em, below left=of lt1] (lv1) {$v$}; 
  \node[ipoint,below right=of lv1] (ls1) {};  
  \node[ipointlabel,below=of ls1] (ls1l) {$b_y$};  
  \path[->] (lv1) edge (lt1)
            (ls1) edge (lv1);
  }
  \begin{pgfonlayer}{background}
  \node[rectangle, fill=black!5, draw=black!30,rounded corners=1ex, fit=(lv1) (ls1) (lt1) (ls1l) (lt1l)] (L) {};  
  \end{pgfonlayer}

  \node[ipoint, right=of J] (rs1) {};  
  \node[ipointlabel, below=of rs1] (rs1l) {${b}_{x}'$};  
  { [node distance=0.5em and 2em]
  \node[ivert, inner sep=0.7em, below right=of rs1] (rv1) {$v'$}; 
  \node[ipoint, below left=of rv1] (rt1) {$$};  
  \node[ipointlabel, below=of rt1] (rt1l) {$b_{y}'$};  
  \path[->] (rv1) edge (rt1)
            (rs1) edge (rv1);
  }
  \begin{pgfonlayer}{background}
  \node[rectangle, fill=black!5, draw=black!30,rounded corners=1ex, fit=(rv1) (rs1) (rt1) (rs1l) (rt1l)] (R) {};  
  \end{pgfonlayer}

  \node[below=of J] (kc) {};  
  { [node distance=0.5em and 1em]
  \node[ivert, inner sep=0.7em, left=of kc] (kv1) {$v$}; 
  \node[ivert, inner sep=0.7em, right=of kc] (kv2) {$v'$}; 
  
  \node[ipoint,above=of kc] (bx) {}; 
  \node[ipoint,below=of kc] (by) {}; 
  \node[ipointlabel,above=of bx] (bxl) {$p_x$};  
  \node[ipointlabel,below=of by] (byl) {$p_y$};  

  \draw[->] (kv1) -- (bx);
  \draw[->] (bx) -- (kv2);
  \draw[->] (kv2) -- (by);
  \draw[->] (by) -- (kv1);
%  \draw[->] (kv1) -- (bx) -- (kv2)
%  (kv2) -- (by) -- (kv1);
%  \path[->] (kv1) edge (bx) edge (kv2)
%            (kv2) edge (by) edge (kv1);
  }

  \begin{pgfonlayer}{background}
  \node[rectangle, fill=black!5, draw=black!30,rounded corners=1ex, fit=(kv1) (kv2) (bxl) (byl)] (M) {};  
  \end{pgfonlayer}

  \path[->,shorten <= 3pt,shorten >= 3pt] 
  (J) edge node[above] {$b$} (L)
  (J) edge node[above] {$b'$} (R)
  (L) edge node[above] {$m$} (M)
  (R) edge node[above] {$m'$} (M);

{ [node distance=0em and 0em]
    \node[above=of M, circle, inner sep=3pt] (pushout) {};
    \draw[line width=1pt] (pushout.west) -- (pushout.north) -- (pushout.east); 
  }

  \end{tikzpicture}
\end{center}
\end{example}

\begin{proposition} \label{prop:plug-pres-hom}
  Plugging respects the equivalence class of homeomorphisms:
  $G \plugw{(K,e_1,e_2)} H \sim G' \plugw{(K,e_1',e_2')} H'$ iff $G' \sim G \text{ and } H' \sim H$, where $e_1'$, $e_2'$ are the embeddings $e_1$ and $e_2$, but considered as maps into $G'$ and $H'$, respectively.
\end{proposition}

This allows any definitions built by plugging graphs together to be
lifted to the equivalence class induced by homeomorphism.

\begin{proposition} \label{prop:gen-merge}
  The most general plugging with respect to matching is $\tensor$:
  \[(H \plugw{p} G \matches K) \Rightarrow (H \tensor G \matches K) \]
\end{proposition}

%*** FIX ABOVE: fixed definition makes this stronger: with repect to embedding. 

%\begin{proposition}
%$\tensor$ is associative and commutative.
%and forms a bi-functor
%  w.r.t. non-empty plugging.
%\end{proposition}

% While plugging composes two graphs, keeping the internal structure, a
% dual concept is that of unification, which composes two graphs but
% merges some of the internal structure. In particular, unification
% describes how two graphs can overlap.

% *** 
% \begin{proposition} 
%   A unifier cannot be constructed by plugging its parts:
%   $$ \plugsss{G}{K}{H} \neq \unifss{G}{H}$$
% \end{proposition}

%\begin{corollary} 
%  Plugging and unification are always different:
%  $$ \plugss{G}{H} \neq \unifss{G}{H} $$
%\end{corollary}

\begin{proposition}
  \label{prop:emb-def-plug} Every open embedding $e : G \injmap M$ defines
  a unique graph $H$ that is all of $M$ except for the image of $e$
  (upto the boundary points). In particular, it identifies a unique
  $p$ such that $G \plugw{p} H = M$.
\end{proposition}

\begin{proposition}
  Every merging $G \mergew{(K,e_1,e_1)} H$ can be written as graphs
  $G'$ and $H'$ that are plugged onto $K$ such that
  $G \mergew{(K,e_1,e_1)} H = G' \plugw{p} K \plugw{q} H'$, 
  $G = G' \plugw{p} K$, 
  $H = K \plugw{q} H'$. 
\end{proposition}

This, combined with plugging respecting homeomorphism
(proposition~\ref{prop:plug-pres-hom}), allows merging (and any other
derived notions) to also respect the homeomorphism. 

\begin{definition}[Subtraction]
  We define the \emph{subtraction} of $G$ from $M$, at $e : G \injmap
  M$, written $M-_e G$, as the unique graph $H$ such that
  $G \plugw{p} H = M$. 
\end{definition}

When the embedding is implicit, we write $M-G$. 
%Given $H = M-_eG$, we
%let $\bar{e} : H \injmap M$ be the embedding of $H$ into $M$ induced
%by the subtraction.

\begin{proposition} 
  Subtracts always embed: $G - H \injmap G$
\end{proposition}

\begin{definition}[Substitution]
  A graph $G$ occurring within $M$ can be substituted, at $e : G
  \injmap M$, for another graph $G'$, when $G$ and $G'$ share the same
  boundary (a subset of which is identified by $P$, $b$ and $b'$). The
  resulting graph, $M'$ is defined according to the following pair of
  pushouts:

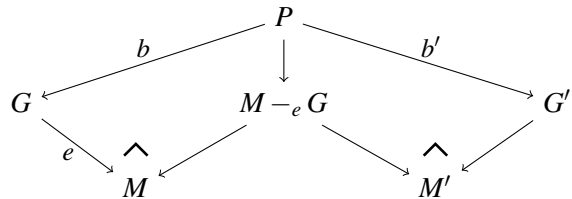
\begin{center}
  \begin{tikzpicture}
  \matrix (m) [row sep=1.5em, column sep=2.5em, text height=1.5ex, text depth=0.25ex] { 
                       &  & \node (top) {$P$};  & & \\
  \node (left) {$G$}; & \node (eq1) {$$}; &  \node (bot) {$M-_{e}G$}; & \node (eq2) {$$}; & \node (right) {$G'$};  \\
  & \node (bleft) {$M$};  &                     & \node (bright) {$M'$}; & \\
  };
  \path[->] (top) edge node [above left,elabel] {$b$} (left)
            (top) edge node [above right,elabel] {$b'$} (right)
            (top) edge node [right,elabel] {$ $} (bot)
            (bot) edge node [above left,elabel] {$ $} (bleft)
            (left) edge node [below left,elabel] {$e$} (bleft)
            (bot) edge node [above right,elabel] {$ $} (bright)
            (right) edge node [below right,elabel] {$ $} (bright)
            ;

{ [node distance=0em and 0em]
    \node[above=of bleft, circle, inner sep=3pt] (pushout) {};
    \draw[line width=1pt] (pushout.west) -- (pushout.north) -- (pushout.east); 
  }
{ [node distance=0em and 0em]
    \node[above=of bright, circle, inner sep=3pt] (pushout) {};
    \draw[line width=1pt] (pushout.west) -- (pushout.north) -- (pushout.east); 
  }

\end{tikzpicture}
\end{center}

\noindent where the left pushout computes the subtraction $M-_{e}G$,
and the right pushout then computes the plugging to form
$(M-_{e}G) \mergew{p} G'$.

%\noindent where $H$ is defined by $K-G_{l}$. The right pushout then
%uses $H$ to define $K'$. 
\end{definition}

Analogously to traditional notation for substitution, we will write
$M'$ as $M[G \mapsto G']_{e}$, and when the embedding is implicit we
write $M[G \mapsto G']$.

% Then $M[R_1 \mapsto L_1]_m$ has the same boundary as $M[L_2\mapsto R_2]_{m'}$.

\begin{proposition}
  Substitution preserves the boundary: $M$ and $M[G \mapsto G']$ have
  the same boundary.
\end{proposition}

\begin{example}[Circles] 
We can now returning to the challenging example introduced at the end
of \S~\ref{sec:elec}. We will rewrite the graph
\begin{tikzpicture}[circuit, node distance=0.5em and 0.7em]
\node (x1) [] {};
\node (g) [andg, right=of x1, inner sep=0.4em] {};
\node (x2) [right=of g] {};
\node (x3) [below=of g] {};
\draw [dline] (g) -- (x2.center) -- (x3.center) -- (x1.center) -- (g);
\end{tikzpicture} 
by 
\begin{tikzpicture}[circuit]
\node (x1) [halfe] {};
\node (g) [andg, right=of x1, inner sep=0.4em] {};
\node (x2) [halfe,right=of g] {};
\draw [dline] (x1) -- (g) -- (x2);
\end{tikzpicture}
=
\begin{tikzpicture}[circuit]
\node (x1) [halfe] {};
\node (x2) [halfe,right=of x1] {};
\draw [dline] (x1) -- (x2);
\end{tikzpicture}. 
to get
\begin{tikzpicture}[circuit, node distance=0.2em and 0.2em]
\node (x1) [] {};
\node (g) [above right=of x1] {};
\node (x2) [below right=of g] {};
\node (x3) [below right=of x1] {};
\draw [dline] (g.center) -- (x2.center) -- (x3.center) -- (x1.center) -- (g.center);
\end{tikzpicture}. 
The pushout construction for this substitution is as follows:

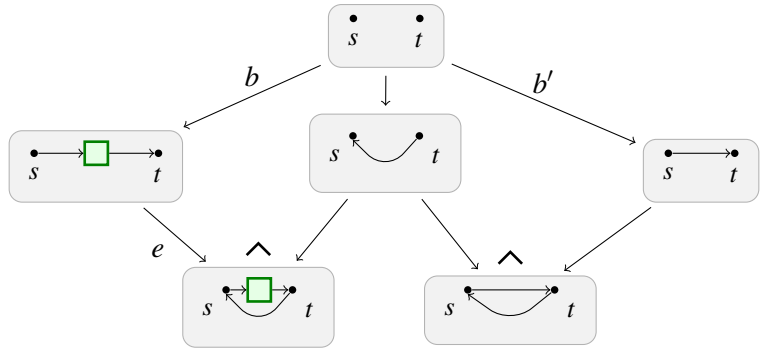
\begin{center}
\begin{tikzpicture}[mcgraph,node distance=5em and 7em]
  %%% Boundary of rule
  { [node distance=0.5em and 2em]
  \node[ipoint] (js1) {};  
  \node[ipointlabel, below=of js1] (js1l) {$s$};  
  \node[ipoint, right=of js1] (js2) {};  
  \node[ipointlabel, below=of js2] (js2l) {$t$};  
  }
  \begin{pgfonlayer}{background}
  \node[rectangle, fill=black!5, draw=black!30,rounded corners=1ex, fit=(js1) (js2) (js2l) (js1l)] (J) {};  
  \end{pgfonlayer}

  %%%% middle part, K1 - L: 
  \node[node distance=2em and 1em, below=of J] (m1c) {};  
  { [node distance=0.5em and 1em]
  \node[inner sep=0.4em,at=(m1c)] (m1x) {}; 
  \node[ipoint, left=of m1x.center] (m1s) {}; 
  \node[ipoint, right=of m1x.center] (m1t) {};   
%  \node[node distance=0.2em and 1em,below=of k1s] (k1s2) {}; 
%  \node[node distance=0.2em and 1em,below=of k1t] (k1t2) {};   
  \node[ipointlabel,below left=of m1s] (m1sl) {$s$};  
  \node[ipointlabel,below right=of m1t] (m1tl) {$t$};  
%  \draw[->] (m1s) -- (m1x);
%  \draw[->] (m1x) -- (m1t);
  \node[below=of m1x] (m1y) {}; 
  \draw[rounded corners=2ex,->] (m1t) -- (m1y.center) -- (m1s.south);
  }
  \begin{pgfonlayer}{background}
  \node[rectangle, fill=black!5, draw=black!30,rounded corners=1ex, fit=(m1x) (m1sl) (m1tl) (m1y)] (M) {};  % (k1s2) (k1t2)
  \end{pgfonlayer}

  %%% L
  \node (lc) [left=of M] {};
  { [node distance=0.5em and 2em]
  \node (lv1) [andg, at=(lc), inner sep=0.4em] {};
  \node (lt1) [ipoint, right=of lv1.center] {};
  \node (ls1) [ipoint, left=of lv1.center] {};
  \node[ipointlabel, below=of ls1] (ls1l) {$s$};  
  \node[ipointlabel, below=of lt1] (lt1l) {$t$};  
  \path[->] (lv1) edge (lt1)
            (ls1) edge (lv1);
  }
  \begin{pgfonlayer}{background}
  \node[rectangle, fill=black!5, draw=black!30,rounded corners=1ex, fit=(lv1) (ls1) (lt1) (ls1l) (lt1l)] (L) {};  
  \end{pgfonlayer}

  %%%% rhs
  \node (rs1) [ipoint, right=of M] {};
  { [node distance=0.5em and 2em]
  \node (rt1) [ipoint, right=of rs1] {};
  \node[ipointlabel, below=of rs1] (rs1l) {$s$};  
  \node[ipointlabel, below=of rt1] (rt1l) {$t$};  
  \path[->] (rs1) edge (rt1);
  }
  \begin{pgfonlayer}{background}
  \node[rectangle, fill=black!5, draw=black!30,rounded corners=1ex, fit=(rs1) (rt1) (rs1l) (rt1l)] (R) {};  
  \end{pgfonlayer}

  %%%% Graph before rewriting K1: 
  \node[node distance=3em and 4em, below left=of M.south] (k1c) {};  
  { [node distance=0.5em and 1em]
  \node[andg, inner sep=0.4em,at=(k1c)] (k1x) {}; 
  \node[ipoint, left=of k1x.center] (k1s) {}; 
  \node[ipoint, right=of k1x.center] (k1t) {};   
%  \node[node distance=0.2em and 1em,below=of k1s] (k1s2) {}; 
%  \node[node distance=0.2em and 1em,below=of k1t] (k1t2) {};   
  \node[ipointlabel,below left=of k1s] (k1sl) {$s$};  
  \node[ipointlabel,below right=of k1t] (k1tl) {$t$};  
  \draw[->] (k1s) -- (k1x);
  \draw[->] (k1x) -- (k1t);
  \node[below=of k1x] (k1y) {}; 
  \draw[rounded corners=2ex,->] (k1t) -- (k1y.center) -- (k1s.south);
  }
  \begin{pgfonlayer}{background}
  \node[rectangle, fill=black!5, draw=black!30,rounded corners=1ex, fit=(k1x) (k1sl) (k1tl) (k1y)] (K1) {};  % (k1s2) (k1t2)
  \end{pgfonlayer}

  %%% K2: K1[L -> R]: the circle
  \node[node distance=3em and 4em, below right=of M.south] (k2c) {};  
  { [node distance=0.5em and 1em]
  \node[ipoint, left=of k2c] (k2s) {}; 
  \node[ipoint, right=of k2c] (k2t) {}; 
  \node[below=of k2c] (k2y) {}; 
  \node[ipointlabel,below left=of k2s] (k2sl) {$s$};  
  \node[ipointlabel,below right=of k2t] (k2tl) {$t$};  
  \draw[->] (k2s) -- (k2t);
  \draw[rounded corners=2ex,->] (k2t) -- (k2y.center) -- (k2s.south);
  }
  \begin{pgfonlayer}{background}
  \node[rectangle, fill=black!5, draw=black!30,rounded corners=1ex, fit=(k2s) (k2t) (k2sl) (k2tl) (k2y)] (K2) {};  
  \end{pgfonlayer}

  %%%%%%% edges between obj of category
  \path[->,shorten <= 3pt,shorten >= 3pt] 
  (J) edge node[above] {$b$} (L)
  (J) edge node[above] {$b'$} (R)
  (J) edge node[above] {} (M)
  (L) edge node[below left] {$e$} (K1)
  (M) edge node[above] {} (K1)
  (M) edge node[above] {} (K2)
  (R) edge node[above] {} (K2);
  %%% pushouts
  { [node distance=0em and 0em]
    \node[above=of K1, circle, inner sep=3pt] (pushout) {};
    \draw[line width=1pt] (pushout.west) -- (pushout.north) -- (pushout.east); 
  }
  { [node distance=0em and 0em]
    \node[above=of K2, circle, inner sep=3pt] (pushout) {};
    \draw[line width=1pt] (pushout.west) -- (pushout.north) -- (pushout.east); 
  }

  \end{tikzpicture}
\end{center}
\end{example}

\begin{definition}
  A \emph{rewrite}, $r$, is a pair of graphs, $(L,R)$, with the
  same boundary.
\end{definition}

We will denote that $(L,R)$ is a rewrite named $r$ by $r : L
\rightarrow R$.

%%% *** note: fixed def allows this to be embedding. 

\begin{definition}
  \emph{Applying a rewrite} $r : L \rightarrow R$ to a graph $G$ at a
  \emph{match} $e : L \hookrightarrow G'$ (for $G' \sim G$) is said to \emph{rewrite} $G$ to
  $H$, when $H = G'[L \mapsto R]_e$. The induced rewrite $G
  \rightarrow H$ is called an \emph{extension} of $r$,
  and named $\exten{r}{e}$.
\end{definition}

%%%%%%%%%%%%%%%%%%%%%%%%%%%%%%%%%%%%%%%%%%%%%%%%%%%%%%%%%%%%%%%%%%%%%%%%%
\section{Graphical Theories}
\label{sec:theories}

We will now develop a categorical description of \emph{graphical
  theories}. This allows the usual concepts from rewriting theory,
such as normalisation, termination, confluence, etc. to be
employed. However, rather than concern ourselves with these familiar
concepts, we will focus on the structure of graphical theories in
terms of how rules can be combined, sequentially and in parallel,
using the underlying graph operations.

\begin{definition}[Sequential Merging of Rewrites]
  Given rewrite rules $r : L \rightarrow R$ and $r' : L' \rightarrow
  R'$, and a merged graph $M := R \mergew{p} L'$, with $p := (K,e,e')$
  and open embeddings $m : R \injmap M$ and $m' : L' \injmap M$
  induced by the pushout, then the \emph{sequential merging} of $r$
  and $r'$ at $p$ defines the new rewrite rule:
  \[
  (\seqcompww{r}{p}{r'}) : (M[R \mapsto L]_m) \rightarrow (M[L' \mapsto R']_{m'})
  \]
\end{definition}

Sequential merging does nothing more than rewrites with extension:

\begin{proposition}[Soundness] \label{prop:soundness}
  if $(\seqcompww{r_1}{p}{r_2}) : G \rightarrow G'$ is a rewrite; then there exists
  a graph $H$, and embeddings $e_1$ and $e_2$ such that $G
  \stackrel{\exten{r_1}{e_1}}{\longrightarrow} H
  \stackrel{\exten{r_2}{e_2}}{\longrightarrow} G'$.
\end{proposition}

While merging provides sequential composition, lifting the concept of
graph plugging to rewrites provides a notion of parallel composition.

\begin{proposition}
  Given rewrite rules $r : L \rightarrow R$ and $r' : L' \rightarrow
  R'$; then $L \plugw{p} L'$ has the same boundary as $R \plugw{p}
  R'$.
\end{proposition}

\begin{proposition}[Completeness]
  if $M \stackrel{\exten{r_1}{e_1}}{\longrightarrow} M_2
  \stackrel{\exten{r_2}{e_2}}{\longrightarrow} M_3$ then there exists
  an $e'$ and $p$ such that $\exten{(\seqcompww{r_1}{p}{r_2})}{e'} : M
  \longrightarrow M_3$
\end{proposition}

Completeness can also be thought of as `compressing' any sequence of
rewrites into a single larger rewrite ($\seqcompww{r_1}{p}{r_2}$) that
does all the steps at once.

\begin{definition}[Parallel composition of rewrites]
  Given rewrites $r : L \rightarrow R$ and $r' : L' \rightarrow R'$,
  the \emph{plugging of rewrites}, also called the \emph{parallel
    composition of rewrites}, is defined as
\[ (r_1 \plugw{p} r_2) :
  (L_1 \plugw{p} L_2) \rightarrow (R_1 \plugw{p} R_2)
  \]
\end{definition}

\begin{proposition} \label{prop:rw-plug-is-a-merge} Plugging is a special case of merging:
  for every plugging of rewrites $p$, the sequential merging produces
  the same rewrite $r_1 \plugw{p} r_2 = \seqcompww{r_1}{q}{r_2}$
\end{proposition}

A special case of plugging rewrites together is the tensor product of
two rewrites: $r \tensor r' := (L \tensor L', R \tensor R')$.

\begin{proposition}
  If $(r_1 \tensor r_2) : G \rightarrow M$ is a rewrite then there
  exists an $H$ and $H'$ such that
\[
 G \stackrel{r_1}{\rightarrow}  H \stackrel{r_2}{\rightarrow} M
\qquad \text{and} \qquad
 G \stackrel{r_2}{\rightarrow}  H' \stackrel{r_1}{\rightarrow} M
 \]
\end{proposition}

\begin{proposition}
  The most general parallel composition of rewrites is the tensor
  product. Given $\exten{(r \plugw{p} r')}{e_1} : G \rightarrow H$
  then there exists $e_2$ such that $\exten{(r \tensor r')}{e_2} : G
  \rightarrow H$
\end{proposition}

We now observe that the extension of a rewrite rule can also be
understood in terms of plugging.

\begin{proposition}
  Given a rewrite rule $r : G \rightarrow G'$, and an extension of it
  $r' : M \rightarrow M'$, then there is an identity rule $\id{H} : H
  \rightarrow H$, such that $r' = r \plugw{p} \id{H}$ for some $p$.
\end{proposition}

%\begin{definition}[Rewrite System]
%  A rewrite system, $S$, is defined by a set of rewrite rules which
%  contains $(G,G)$ for all $G$.
%\end{definition}
 
%\begin{definition}[Rewrite expression]
%A rewrite expression, formed from a set of rewrites $R$, is a
%sequential composition of rewrites 
%\end{definition}

\begin{definition}[Graphical Theory]
  Given $\Gamma := (S,R)$, where $S$ is a \emph{generating set of
    graphs}, closed under plugging, and a $R$ is a \emph{generating
    set of rewrites}, closed under extension and formed from $S$; the
  \emph{graphical theory}, $\catGThy(\Gamma)$ is the category with $S$
  as the objects and arrows formed by formal sequential compositions
  of rewrites in $R$. Each arrow is defined by a particular
  composition. The generating set $R$ is also called the \emph{rewrite
    rules} of the theory, while compositions are called \emph{rewrite
    sequences}.
\end{definition}

\begin{proposition}
  Every graphical theory is a monoidal category, with tensor product
  on objects inherited from \catOGraph, and on rewrites as the special
  case of plugging described above.
\end{proposition}

A model of computation can now be characterised by a particular
graphical theory. The generating graphs describe the objects of
interest and axioms are the generating rewrites rules which describe
the model's interesting structural dynamics. For instance, the example
introduced in \S\ref{sec:elec} is a graphical theory.

A direct result of proposition\ref{prop:rw-plug-is-a-merge} is that
graphical theories also have parallel and tensor composition. A
consequence of soundness (proposition \ref{prop:soundness}) for
graphical theories is that, given an initial set of rules, new rules
in the theory may be safely derived by sequential merging of existing
ones; these new rules will also be in the graphical theory. This gives
an algorithm to derive new rules from existing ones which are treated
as axioms of a computational model.

%\begin{proposition}[Admissibility]
%  Given any two rewrites $r_1$ and $r_2$ in a graphical theory,
%  $\Gamma$, every sequential merging $\seqcompww{r_1}{p}{r_2}$ is in
%  $\Gamma$.
%\end{proposition}

%\begin{proposition}
%  Compositions in a rewrite system are admisible: given $r_1, r_2 \in
%  S$, $S = S' \cup \{(r_1 ; r_2)\}$ and $G \rightarrow^*_S H$, then $G
%  \rightarrow^*_{S'} H$
%\end{proposition}

%\begin{proposition}
%  An arrow in a graphical theory, $r : G \rightarrow M$, is either a
%  generating rewrite rule of the theory or can be expressed as the
%  merging of generating rewrites.
%\end{proposition}

%\hspace{0.5cm} or \hspace{0.5cm} 
%$(\plugss{r_1}{r_2}) : G \rightarrow M$
%\end{center}

\section{Conclusions and Further Work}
\label{sec:concl}

We have formalised a compositional account of graphs which represent
computational processes. These graphs have an interface made of
half-edges that enter or leave the graph. Methods to support graphical
rewriting have been described, and it has been shown how graphical
rewriting rules can themselves be composed. The construction is by an
richer intermediate notion of graphs with edges-points. These more
exotic structures provide the needed structure for composition by
pushouts. In particular, they allow rewriting to preserve a graphs
interface, even in the presence of cycles.

This foundation allows a variety of techniques from rewriting, such as
Knuth-Bendix completion~\cite{knuth-bendix}, to be lifted to reasoning
about computational graphs. Another area of further work is to
extending this formalism to include bang-boxes, as introduced
in~\cite{2009:DixonDuncan:AMAI}, as well as other kind of iterative
and recursive structure. We have started to study the categorical
structure of graphical theories, but there is a lot more to be
considered, such as the relationship between graphical theories and
the category of open graphs. The exact of the relationships between
open graphs, topological analogies, and other graphical formalisms is
also important future research.

{\small{\paragraph{Acknowledgements}
This research was funded by EPSRC grants EPE/005713/1 and EP/E04006/1,
and by a Clarendon Studentship. Thanks also to Jeff Egger and Rod
Burstall for their helpful discussions.}

\bibliographystyle{plain}
\bibliography{bibfile}

\end{document}